\input{aipcheck}

\documentclass[
    ,final            
  ]
  {aipproc}

\layoutstyle{6x9}

\def \agile {{\it AGILE} }
\def \egret {{\it EGRET} }
\def \glast {{\it GLAST} }

\def \gray {$\gamma$-ray }

\def \apj {ApJ}
\def \apjs {ApJS}

\def \mnras {MNRAS}

\begin{document}

\title{The Duty-cycle of Gamma-ray Blazars: a New Approach, New Results}

\author{S. Vercellone}{
  address={IASF-CNR Sezione di Milano, Via Bassini 15, 20133 Milano - Italy}
}

\author{S. Soldi}{
  address={ISDC,
  Chemin d'{\`E}cogia 16, 1290 Versoix - Switzerland}
}

\author{A.W. Chen}{
  address={IASF-CNR Sezione di Milano, Via Bassini 15, 20133 Milano - Italy}
  ,altaddress={CIFS, Viale Settimio Severo 63, Torino - Italy}
}

\author{M. Tavani}{
  address={IASF-CNR Sede di Roma, Via del Fosso del Cavaliere 101, Roma - Italy}
  ,altaddress={CIFS, Viale Settimio Severo 63, Torino - Italy}
}

\begin{abstract}
We study several properties of blazars detected in the gamma-ray energy range
by comparing the \egret sources with a sample of radio blazars
which can be considered possible gamma-ray candidates. We define three
classes: non-gamma-ray blazars, blazars with quasi-steady gamma-ray emission,
and gamma-ray blazars with substantial activity level. 
By combining the information of detected and candidate AGNs, we characterise
the blazar activity, including the discovery of a region of consistency between 
the gamma-ray flaring duty-cycle and the recurrence time between flares.
We also find a possible relation between the activity index of FSRQs and their
black hole mass. 

\end{abstract}

\maketitle


\section{Introduction}
Among Active Galactic Nuclei (AGN), {\it blazars}
show strong flux variability at almost all frequencies of the
spectral energy distribution (SED). The \egret instrument  on-board of
CGRO detected, above 30~MeV, blazars as a {\em class} of \gray 
sources \cite{Hartman99}, identifying 67 objects, and detecting 27 candidates.
Gamma-ray blazars are characterised by high variability on different 
time--scales, from one day (e.g. PKS~1622-297) to one month 
(e.g. PKS~0208-512).
This large spread in time variability and the sparse coverage obtained by
\egret make it difficult to quantify parameters such as the {\it duty--cycle}
(i.e. the fraction of time spent in a flaring state)
and/or the characteristic time--scale  of \gray activity for this class
of sources. 

A preliminary result was obtained by \cite{Stecker96},
based on the Second \egret Catalogue assuming that the link between
the radio and the \gray emission can be used to derive the \gray luminosity 
function. Under these assumptions, they found that a duty--cycle
$\zeta = 0.03$ is consistent with the data. One weakness
of this analysis is that the radio-gamma relationship is highly
uncertain.

The aim of this Paper is to summarize a recent result
on the estimate of the blazar duty--cycle and \gray variability 
\cite{Vercellone04}.

\section{The Blazar \gray variability}
In order to investigate the blazar \gray duty--cycle, we analysed 
the Third \egret Catalogue (3EG) AGN sample, looking for recurrent activity.
For each source we considered the number of viewing
periods (VPs) during which it was close enough to the pointing direction
($<40^{\circ}$, the FOV radius) that its exposure for that
VP was greater than zero.
For each source we calculated the {\it Exposure} ($EXP$), defined as
the sum of the exposures during each single viewing period.
%
\begin{figure}[!ht]
  \includegraphics[height=.4\textheight]{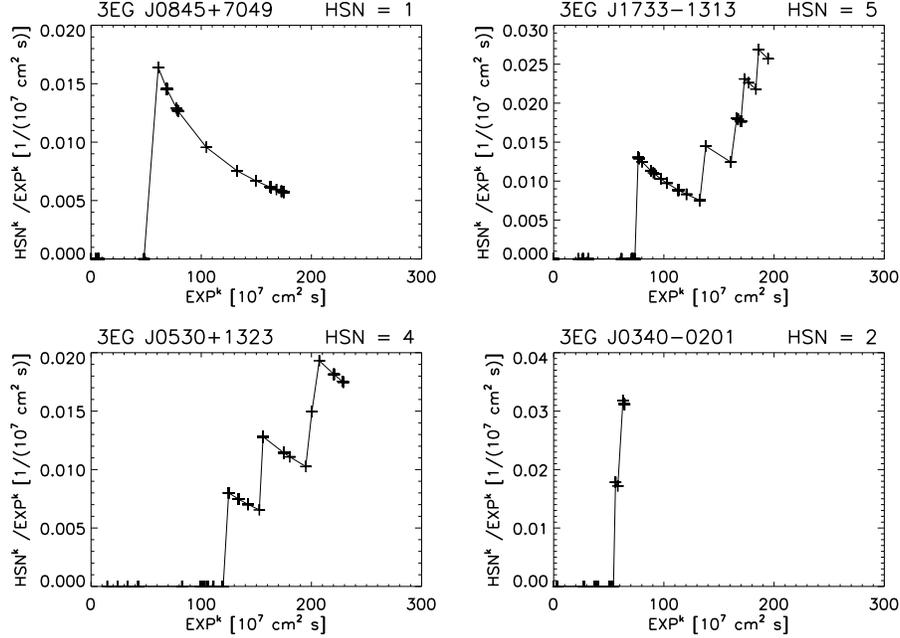}
  \caption{Four different examples of blazar \gray activity as a function
      of exposure. {\it Top-left panel:} 3EG~J0845$+$7049 with $HSN=1$
      despite its high exposure; {\it Top-right panel:}
      3EG~J1733$-$1313 which, for a similar exposure, has $HSN=5$;
      {\it Bottom-left panel:} 3EG~J0530$+$1323 which showed a possible periodicity 
      in its high-state status, and {\it Bottom-right panel: } 3EG~J0340$-$0201 which 
      has a very low exposure ($\sim 64 \times 10^{7}$ cm$^{2}$s), but showed
      2 high-states in a few observations.
      Crosses show the value of the discrete $\psi$ for each VP containing the source.}
    \label{multi-fom}
\end{figure}
%

In order to estimate the \gray activity level, we computed the number of times
that a source was in a {\it high-state} ($HSN$) i.e. when the flux of the
$i$-th VP was greater than a given threshold.
We establish that
a source is in a high-state when 1) the flux of the $i$-th VP is in excess of
1.5 times the mean flux of the source, and 2) the 1-$\sigma$ uncertainty of the 
measurement at the high state is less than the deviation of the measurement 
from the mean.
The mean flux is defined as the 
weighted mean of all the detections and upper limits, assuming the 
weight $w(\sigma)=1/\sigma$, where $\sigma$ is the uncertainty of the
measurement for detections, while $\sigma = UL$ for the upper limits.
%
%
%
If two high-states were less than 2 weeks apart (the typical duration
of an \egret pointing), we gathered them into a single high-state
coincident with the first one.
%

%
Given $EXP$ and $HSN$, we computed an {\it activity index} ($\psi$) of 
each AGN as follows:
\begin{equation}
  \psi_{\rm src} = HSN_{\rm src} \times [EXP_{\rm src}]^{-1}\,\,\,
  [{\rm cm}^{-2}{\rm s}^{-1}]\,.
\end{equation}
This index provides an estimate of blazar \gray activity, weighted by the
exposure. Sources with high $\psi$ have a \gray
activity which is highly variable. Sources with $\psi = 0$
do not show significant flux variability.

We analysed the behaviour of each single source having $\psi \neq 0$
during its time history. Fig.~\ref{multi-fom} shows the discrete $\psi$ 
for some \egret sources as a function of $EXP$.
%
We show four different examples of activity behaviour: 1) 3EG~J0845$+$7049,
with a single $HSN=1$ despite its large exposure; 2) 3EG~J1733$-$1313,
which, for a similar exposure, has $HSN=5$; 3) 3EG~J0530$+$1323,
showing a possible periodicity in its high-state status, and 4)
3EG~J0340$-$0201, which has a very low exposure ($\sim 64 \times 10^{7}$ cm$^{2}$s),
but showed 2 high-states in a few observations.

Sources that have $HSN=1$ (e.g., 3EG~J0845$+$7049) show a rise 
of $\psi_{src}^{k}$ followed by a decay after reaching a single high state,
while sources that have high $HSN$ (e.g. 3EG~J0530$+$1323) show a saw-tooth
pattern, corresponding to alternately increasing/decreasing of $\psi_{\rm src}^{k}$.

We distinguish sources for which the activity level 
$\psi = 0$ (population {\it A}, $36$ per cent), and the ones having
$\psi \neq 0$ (population {\it B}, $64$ per cent).
A remarkable difference appears in the two populations when we compare the
ratio between flat-spectrum radio quasars (FSRQs) and BL-Lac objects
(BLs). Population A is almost equally divided between FSRQs and BLs 
($\sim$60 and $\sim$40 per cent, respectively), while
Population B is dominated by FSRQs ($\sim 80$ per cent).
FSRQs are generally more variable and more luminous in the \gray
energy band than BLs, which could explain the higher fraction of these
sources in Population B.
A possible correlation is found between the activity level and the black hole
mass, with a probability $P_{\rm rand}$ of a randomly distributed 
sample of objects on the order of $\sim 10^{-2}$.
We suggest that the
use of the $\psi$ statistic is more robust than the use of the $HSN$,
since the latter is not weighted by the \egret coverage.
%

\section{The blazar gamma-ray duty--cycle} \label{act_frac}
\begin{figure}[!ht]
  \includegraphics[height=.4\textheight]{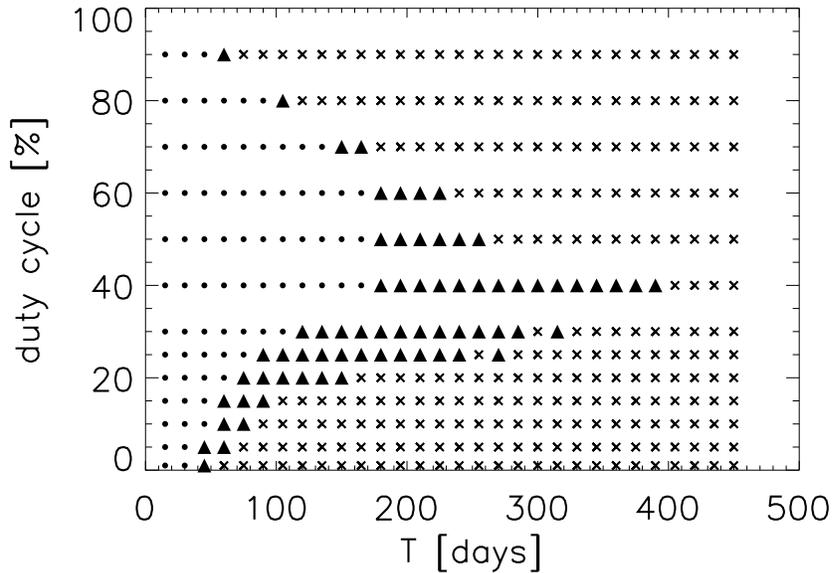}
  \caption{Mean ratio $R_{\rm AB}={\rm \# Pop.A/(\# Pop.A+\# Pop.B)}$ from
    simulations of \egret-detected blazars only.}
  \label{popab_005}
\end{figure}

We develop a simplified
model of \egret blazar activity that yields rough, but quantifiable, estimate
of physical parameters of interest.
In this simplified model, we make several assumptions:
\begin{enumerate}
\item that all of the \egret blazars exhibit the same basic behaviour.  While
observationally unjustified, this assumption is necessary to model
the average behaviour. 
\item that only those blazars detected by \egret are part of the population of
interest. 
\item that the behaviour of all \egret blazars can be characterised by a
simple model with only two free parameters, the duty--cycle, $\chi$, and the 
characteristic time--scale, $T$.  Each blazar will spend a period of time whose
average length is $T$ at a low flux level ({\it off}) before emitting a 
\gray flare of duration $\tau$ at a high flux level ({\it on}), then 
returning to a low level.  The duty--cycle is then defined as the 
fraction of time spent in the {\it on} state:
\begin{equation}
  \chi = \frac{\tau}{\tau + T}\;.
\end{equation}
Note that all blazars have the same characteristic time--scale $T$, but
the duration of each individual quiescent periods is drawn from a Poisson
distribution with mean $T$.
\end{enumerate}
%
We explored the parameter space of $(\chi,T)$.  For each pair of $\chi$ and
$T$ under consideration, we generated 100 sets of 67 simulated light curves
covering the entire time interval of \egret Cycle 1 to 4 (1620 days),
one for each blazar. The time sequence of {\it on-off} states was determined
by drawing the durations of the {\it off} states from a Poisson distribution 
with mean $T$. The durations of the {\it on} states were fixed at $\tau$.  
To obtain the observed fluxes, we used a bootstrap procedure.  Each light
curve was compared with the observation history of the corresponding AGN.
If a given viewing period (VP) coincided with an {\it on} state, the observed
flux was randomly drawn from the distribution of all fluxes {\it detected} from
all \egret blazars during a single VP. Otherwise, the observed flux was 
randomly drawn from the distribution of all {\it upper limits} 
(i.e. non-detections) at \egret blazar positions during a single VP. 
For each AGN the flux values assigned to each VP were used to calculate its 
$HSN$ (and therefore its categorisation as Population A or B) in the same 
manner as the actual \egret blazar $HSN$. 
The result is 100 sets of simulated blazar light curves, each 
with a ratio $R_{\rm AB}={\rm \# Pop.A/(\# Pop.A+\# Pop.B)}$.  Thus
for each pair of values of $(\chi,T)$, we have obtained a distribution of
100 simulated values of $R_{\rm AB}$, each with a mean and variance. Therefore,
we are able to define a confidence region of $(\chi,T)$ pairs whose mean
values of $R_{\rm AB}$ are within one standard deviation of the observed
value $R_{\rm AB} = 24/67 = 0.36$. For values of $R_{\rm AB}$ close to
the observed value, the simulations give a standard deviation of $0.05$.
We ran 100 simulations each for 390 $(\chi,T)$ pairs, ranging 
from $0<T<450$~days and from $0<\chi<1$.  Fig.~\ref{popab_005} 
shows the results.
%
%
%
%
Points in  Fig.~\ref{popab_005} represent the mean value of $R_{\rm AB}$ for 
each value of $T$ and $\chi$. The filled triangles represent
those ratio values between $0.31 \le R_{\rm AB} \le 0.41$, small filled
circles represent $0.0 \le R_{\rm AB} < 0.31$, while crosses represent 
$0.41 < R_{\rm AB} \le 1.0$.  Taking into account our simplifying assumptions,
including the restriction that the observed \egret blazars form the entire
population of \gray emitting blazars, the only region consistent with the
observations is the region where $0.31 \le R_{\rm AB} \le 0.41$.  The
region with high $R_{\rm AB}$ produces too few flaring sources, while
the region with low $R_{\rm AB}$ produces too many.  This implies that the
true characteristic time--scale $T$ lies between 50 and 400 days, the
duty--cycle $\chi$ becoming more extreme for lower values of $T$.
Note that for the Stecker \& Salamon duty--cycle of 0.03 the characteristic
time--scale $T$ would be on the order of 50 days.

If we should take into account a wider sample of \gray blazar, by including
radio blazars emitting at a level of 250~mJy, we would obtain that the 
low $R_{\rm AB}$ region to remain excluded, because adding more sources is 
unlikely to reduce the absolute number of Population B sources predicted, 
which is higher than the number observed. 
On the other hand, the $(\chi,T)$ pairs in the high $R_{\rm AB}$ region
may now be consistent with the \egret data.
Therefore, the region in $(\chi,T)$ space where formerly the $R_{\rm AB}$ was 
consistent with the observed value now forms a lower
bound to the possible values of $T$ for a given value of $\chi$.

\section{The Black Hole Mass -- Activity Relationship} \label{act_mbh}
Recently, \cite{Fan04} 
estimated the black hole mass for 36 \egret AGNs from their optical continuum 
luminosity; they claim a positive correlation (significance level 96.6 per cent) between
the maximum \gray source luminosity and the black hole mass.

We look for a possible correlation between the \gray maximum
luminosity, $L_{\gamma}^{\rm max}$, and the black hole mass, $M_{\rm BH}$, 
obtained by adopting the black hole masses reported in \cite{Woo02} for
all the \egret blazars except for 3C~273, whose $M_{\rm BH}$ was taken from
\cite{Kaspi00}. 
The correlation is absent when considering
the whole sample (Spearman statistic: $\rho = 0.44$, and $P = 0.15$),
and becomes marginal if we take into account the FSRQs only 
($\rho = 0.60$, and $P=0.08$).
The relation between luminosity and black hole mass is, approximately,
$L_{\gamma} \propto M_{\rm BH}^{0.9}$ for FSRQs, consistent with the predicted 
relation by \cite{Dermer95}, although their assumptions
(\gray emission is isotropic radiation produced by the steady and Eddington--limited
accretion without considering the beaming effect) do not agree with
recent results on \gray blazars \cite{Jorstad01a}.
%
\begin{figure}[!ht]
  \includegraphics[height=.4\textheight]{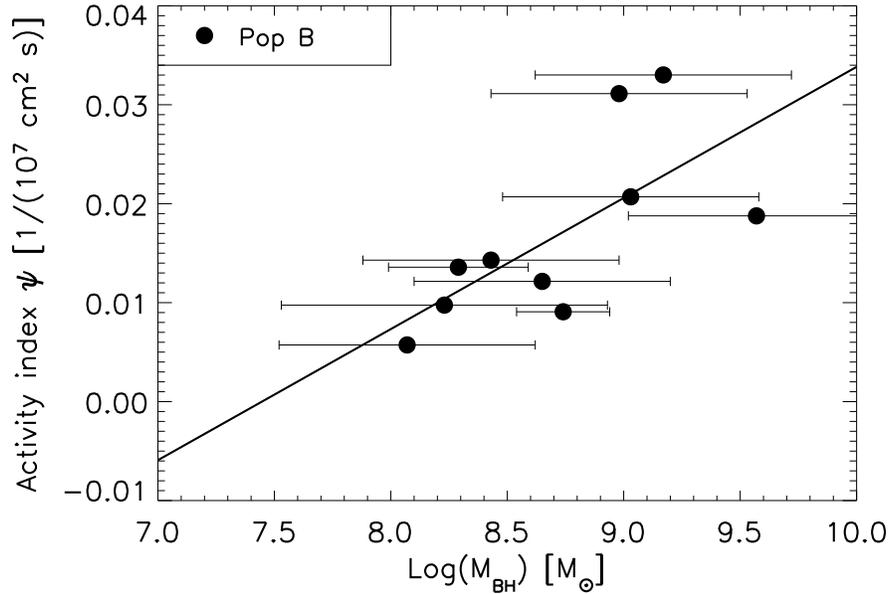}
  \caption{\gray activity ($\psi$) vs. black hole mass ($M_{\rm BH}$) for
    \egret Pop. B blazars. A marginal correlation is present
    ($\rho=0.76$ and $P=0.01$, Spearman statistic). The solid line represents
    the best fit line, $\log (\psi)=0.013\times \log ({\rm M_{\rm BH}})-0.099$.
    Horizontal bars represent the typical black hole 
    mass estimate uncertainty as reported by \cite{Vestergaard04}.}
  \label{act_vs_mbh}
\end{figure}
%
%
%
%
When analysing the possible correlation between the activity level, $\psi$, 
and the maximum \gray luminosity, we find no correlation ($\rho= 0.14$, 
and $P = 0.39)$.
We found a possible correlation between the activity index
$\psi$ and $M_{\rm BH}$.
Fig.~\ref{act_vs_mbh} shows our results based on ten blazars with
non-zero $\psi$ and the black hole mass estimation.
The Spearman correlation parameters turn out to be 
$\rho=0.76$ and $P = 0.011$. 
This possible correlation seems to indicate that more massive 
BHs can induce
higher \gray activity, although our sample is too small to draw a firm
conclusion. More reliable BH mass estimates of a larger blazar sample
will be crucial in verifying the existence of this correlation.

\section{Conclusions}
The results presented here summarize a more detailed analysis presented in
\cite{Vercellone04} where we discuss the implications on the duty--cycle
estimate of a sample of candidate \gray AGNs extracted from recently published
radio catalogues. A firm estimate of the duty--cycle will be a scientific goal 
of future \gray missions such as \agile and \glast. Their wide FOV ($\sim60^{\circ}$)
will allow monitoring of a large number of AGNs for each pointing on long time--scales.
For the time being, an extensive black hole
mass estimation by means of optical spectroscopy on FSRQs and BL~Lac objects will
allow us to better understand the relation between the  central engine 
and the \gray activity.







\end{document}